\documentclass[useAMS,usenatbib,twocolumn,psfig,fleqn,amstex,rotating]{mn2e}
\usepackage{graphicx}
\usepackage{color}

\newif\ifAMStwofonts
\AMStwofontstrue

\def\caM{{\cal M}}

\def\gsim{~\rlap{$>$}{\lower 1.0ex\hbox{$\sim$}}}

\def\xm{x_{_{\rm M}}}

\def\ltsim{\lower.5ex\hbox{$\; \buildrel < \over \sim \;$}}
\def\gtsim{\lower.5ex\hbox{$\; \buildrel > \over \sim \;$}}
\def\ltsim{\lower.5ex\hbox{$\; \buildrel < \over \sim \;$}}
\def\gtsim{\lower.5ex\hbox{$\; \buildrel > \over \sim \;$}}

\def\vx{{\bf x}}

\def\vr{{\bf r}}

\def\vv{{\bf v}}

\def\pa{\partial}
 
\def\vg{{\bf g}}

\def\dd{\,{\rm d}}

\def\gN{g_{_{\rm N}}}
\def\gM{g_{_{\rm M}}}

\def\prd{{Phys. Rev. D}}
\def\physrep{{Phys.~Rep.}}   
\def\etal{et  al.\ }

\title[Cluster formation in MOND]{Modeling the formation of galaxy
  clusters in MOND}

\author[A. Nusser \& E. Pointecouteau ] { Adi Nusser$^{1,2}$\thanks{E-mail: adi@physics.technion.ac.il} and Etienne Pointecouteau$^{2}$  \\\\
  $^{1}$Physics Department- Technion, Haifa 32000, Israel\\
  $^{2}$Astrophysics, Oxford University, Keble Road, Oxford OX1 3HR, UK}

\begin{document}
\date{;}
\pagerange{\pageref{firstpage}--\pageref{lastpage}} \pubyear{2002}

\maketitle

\label{firstpage}

\begin{abstract}
  
  We use a one dimensional hydrodynamical code to study the evolution
  of spherically symmetric perturbations in the framework of Modified
  Newtonian Dynamics (MOND).  The code evolves spherical gaseous
  shells in an expanding Universe by employing a MOND-type
  relationship between the fluctuations in the density field and the
  gravitational force, $g$.  We focus on the evolution of initial
  density perturbations of the form $\delta_{i}\sim r_{i}^{-s}$ for
  $0<s<3$. A shell is initially cold and remains so until it
  encounters the shock formed by the earlier collapse of shells nearer
  to the centre.  During the early epochs $g$ is sufficiently large
  and shells move according to Newtonian gravity.  As the physical
  size of the perturbation increases with time, $g$ gets smaller and
  the evolution eventually becomes MOND dominated.  However, the
  density in the inner collapsed regions is large enough that they
  re-enter the Newtonian regime.  The evolved gas temperature and
  density profiles tend to a universal form that is independent of the
  the slope, $s$, and of the initial amplitude. An analytic
  explanation of this intriguing result is offered. Over a wide range
  of scales, the temperature, density and entropy profiles in the
  simulations, depend on radius roughly like $r^{0.5}$, $r^{-1.5}$ and
  $r^{1.5}$, respectively.  We compare our results with XMM-{\it
    Newton} and {\it Chandra} observations of clusters.  The
  temperature profiles of 16 observed clusters are either flat or show
  a mild decrease at $R\gtsim 200\rm kpc$.  MOND profiles show a
  significant increase that cannot reconciled with the data.  Our
  simulated MOND clusters are substantially denser than the observed
  clusters.  It remains to be seen whether these difficulties persist
  in three-dimensional hydrodynamical simulations with generic initial
  conditions.
     
\end{abstract}
 
                           
\section{introduction}

In the standard cosmological paradigm the Universe is predominantly
filled with dark matter (DM).  The success of this paradigm is
indisputable. It explains a multitude of observational data over a
wide range of scales, from rotation curves of local galaxies to the
anisotropies of the cosmic microwave background (CMB) at high
redshifts.

Despite this success, several fundamental problems remain to be
resolved. Although the best evidence for DM is the flattening of
rotation curves of many spiral galaxies, the scenario does not explain
the shape of these curves in the inner regions of some galaxies (e.g
de Blok et al. 2001, Hayashi et~al. 2004). The DM scenario predicts
specific correlations between the internal properties of clusters of
galaxies (e.g. the mass vs temperature, entropy vs temperature). While
such correlations are observed, their parameters do not agree with the
model predictions.  There are various puzzles related to the
evolution of the galaxy population as well.  All these problems and
several others can probably be attributed to physical processes
unrelated to the dark matter.  But they do allow us some leeway in
exploring other alternatives to the DM scenario.  One possibility is
to superimpose additional forces acting only in the dark sector (e.g.
Gradwohl \& Frieman J.A. 1992; Farrar \& Peebles 2004; Sealfon et~al.
2005; Shirata et~al. 2005). These forces can work in the direction of
boosting the clustering on small scales, leaving the larger scales
intact.  Several desirable features follow from this scenario (e.g.
Nusser, Gubser \& Peebles 2005).

Another approach is to modify Newton's $g_{_N}\sim 1/({\rm
  distance})^2 $ law for the force of gravity, relinquishing the DM
scenario altogether.  This approach was originally introduced to
account for the flattening of rotation curves without invoking dark
matter (Milgrom 1983; Bekenstein \& Milgrom 1984).  After all, the DM
particle has yet to be discovered (Bertone \etal\ 2004) and we lack an
experimental verification of Newton's laws at low accelerations.
The essence of Modified Newtonian Dynamics (MOND) is to replace
Newton's law of gravity at sufficiently low accelerations
$g_{_N}<g_{_0} $\footnote{In MOND literature $a_{_0}$ is often used
  instead of $g_{_0}$.}  by $g\sim \sqrt{g_{_0} g_{_N}}$ where
$g_{_0}\approx 1.2 \times 10^{-8}{\mathrm cm \; s^{-2}}$ is found to
give the best results in fitting the rotation curves.

Recently a few attempts have been made to confront MOND with
observations of the large scale structure (LSS) in the Universe
(McGaugh 1999, Sanders 2001, Nusser 2002, Knebe \& Gibson 2004,
McGaugh 2004).  When MOND is applied to a uniform background it
predicts the collapse of any finite region in the Universe regardless
of the mean density in that region (e.g., Felten 1984, Sanders 1998).
To solve this problem Sanders (2001) proposed a two-field Lagrangian
based theory of MOND in which the Friedmann-Robertson-Walker (FRW)
background cosmology remains intact in the absence of fluctuations. He
argued that this theory leads to LSS resembling
Newtonian dynamics with CDM-like initial conditions.  More recently
Bekenstein (2004) proposed a general relativistic version of MOND in
which the behaviour of cosmological background is very close to a FRW.
Nusser (2002) employed the ``Jeans swindle'' (Binney \& Tremain 1987)
to write a MOND type relation between the fluctuations in the density
and the gravitational force field.  The relation can be derived from a
Lagrangian and is equivalent to Sanders' two-field theory in the limit
of small coupling (Sanders 1998). And it seems that a similar relation
can be derived from Bekenstein's (2004) theory as well.  Nusser (2002)
then implemented this relation in a collisionless N-body code to
simulate the evolution of LSS under MOND.
This work showed that MOND, albeit with $g_{0}$ smaller than the
standard value inferred from rotation curves, produces LSS similar to
that seen in simulations of viable variants of the Cold Dark Matter
(CDM) scenario.  Knebe \& Gibson (2004) used high resolution
simulations of collisioness particles to study the LSS and ``halo''
profiles in MOND.  According to these authors, MOND leads to
reasonable predictions for the LSS even for the standard value of
$g_{0}$.  Knebe \& Gibson also examined the properties of halos and
concluded that density profiles in MOND have similar shape to the
profiles seen in Newtonian simulations.

A MONDian Universe is dominated by ordinary baryonic matter made
mainly of primordial hydrogen and helium.  Baryons in cosmological
systems have a relatively short mean free path, and should be treated
as a hydrodynamical fluid.  This fluid can collapse under its
self-gravity, shock-heat during the collapse, cool, and form stars.
Reliable three dimensional simulations of these processes are
difficult and very CPU demanding, even in Newtonian dynamics where the
Poisson equation can be solved efficiently.  The problem is greatly
simplified if one restricts the analysis to symmetric perturbations.
Stachniewicz \& Kutschera (2004) adopted this approach to the collapse
of low mass objects at high redshifts using a spherically symmetric
hydrodynamical code with MONDian gravity.  In this
  paper we address the formation of clusters of galaxies in MOND
using hydrodynamical simulations of spherically symmetric
perturbations having initial power law profiles.  The perturbations
are evolved using a one dimensional Lagrangian hydrodynamical code.
Cooling and heat conduction are not included in this code.  These
processes are important in galaxies and small groups of galaxies.
However, they are likely to play a minor role in governing the
evolution of the gaseous intercluster medium (ICM) in the outer
regions of massive clusters.  Several works have confronted MOND with
observations of clusters (e.g. The \& White 1988; Aguirre et~al.
2001). Aguirre et~al. (2001) assumed the hydrostatic equilibrium to
derive MOND temperature profiles from X-ray data of three nearby
clusters.  They concluded that MOND predictions for the temperature
profiles disagree with observations.  Our approach here is to derive
cluster properties resulting from the dynamical evolution of initial
perturbations. This approach will enable us to make detailed
predictions independently of the observations, and to test several key
assumptions such as that of the hydrostatic equilibrium.  Therefore, we
expect it to provide stringent constraints on MOND from the
observational data.

The remainder of the paper is organised as follows.  The notation, the
equations of motion, and the initial conditions are described in
\S\ref{sec:eom}. An analytic treatment of the evolution when the
perturbations are small is given in \S\ref{sec:small}.  The numerical
model is outlined in \S\ref{sec:nmodel}.  In \S\ref{sec:tests}, tests
using known self-similar solutions are presented. The results for MOND
are described \S\ref{sec:results}. In \S\ref{sec:obs} our results are
confronted with recent observations. A final summary and discussion
are presented in \S\ref{sec:sum}.

\section{The modified cosmological equations of motion}
\label{sec:eom}
The background FRW cosmology is described by the scale factor $a(t)$
normalised to unity at the present, the Hubble function $H(t)=\dot a
/a$, and the total mean background matter density $\bar \rho_{tot}
=\bar \rho_{\rm dm}+\rho_{\rm b}$, where $\bar \rho_{\rm dm}$ and
$\rho_{\rm b}$ are the mean densities of the dark and baryonic matter,
respectively.  In MOND the Universe is made of baryonic matter and so
we take ${\bar \rho}_{\rm dm}=0$. Further, we assume a vanishing
cosmological constant since otherwise the cosmic age would be
unrealistically long if $\rho_{\rm b}$ is fixed by nucleosynthesis.
We also define $\Omega_{\rm b}= \bar \rho_{b}/\rho_c$, where
$\rho_c=3H^2/(8\pi G)$ is the critical density.  These cosmological
quantities are related by Einstein equations of general relativity.
Let $\vr $ and $\vx=\vr/a$ denote, respectively, physical and comoving
coordinates. The fluctuations over the uniform background in the
matter distribution are described by the comoving peculiar
velocity $\vv=\dd \vx/\dd t$ of a patch of matter, the density
contrast $\delta(\vx) =\rho(\vx) /\bar \rho-1$, where $\rho(\vx)$ is
the local density, and the fluctuations in the gravitational force
field, $\vg$.  Further, the thermal state of the gas is described by
the pressure, $P$ and the internal energy, $u$, which are assumed to
be related to the density by a perfect equation of state $P=(\gamma
-1)\rho u$, where $\gamma=5/3$. Here, we find it convenient to work
with $p=P/{\bar \rho}$, instead of the physical pressure $P$.  The
Newtonian equations of motion governing the evolution of a spherical
perturbation are: the continuity equation
\begin{equation}
\frac{\pa \delta}{\pa t }+\frac{1}{x^{2}}\partial_x [x^{2}(1+\delta) v] =0 \; ,
\label{cont}
\end{equation}
the Euler equation of motion,
\begin{equation}
\frac{\dd v }{\dd t }+2H v =\frac{1}{a} g -\frac{1}{a^{2}}\frac{\partial_{x} p}{1+\delta}\; ,
\label{euler}
\end{equation}
 the energy equation,
\begin{equation}
\frac{\dd u}{\dd t}=\frac{p}{(1+\delta)^{2}}\frac{\dd \delta}{\dd t}-
3H u \; .
\end{equation}
where the ``pressure'', $P$, and internal energy, $u$,  are related by the 
equation of state, 
\begin{equation}
P=(\gamma-1)(1+\delta)u \; . 
\end{equation}
The final equation of motion that is needed
is a relation between $g$ and the density, $\delta$.  In the Newtonian
theory $g=g_{_{\rm N}}$ where $g_{_{\rm N}}$ satisfies the Poisson
equation,
\begin{equation}
\frac{1}{x^{2}}\partial_x (x^{2}g_{_{\rm N}}) =
-4\pi G a \bar \rho_{\rm b} \delta=
-\frac{3}{2}a \Omega_{\rm b} H^2 \delta\; .
\label{poisson}
\end{equation}
In MOND $g=\gM$ where $\gM$ is related to $\gN$ by $\gN=\gM \mu(\gM
/g_{0})$ with $\mu(y)=y (1+y^{2})^{1/2}$ (Milgrom 1983, Bekenstein
2004). This relation yields
\begin{equation}
\gM=\gN\sqrt{\frac{1}{2}+\frac{1}{2}
\sqrt{1+\left(\frac{2g_{0}}{\gN}\right)^{2}  }} \; .
\label{eq:gm}
\end{equation}

Neglecting thermal effects, the linear Newtonian theory implies that
$|g|\gg g_{0}$ at sufficiently early times (see \S\ref{sec:small}).
But $|g|$ decreases with time and MOND eventually takes over the
evolution of the perturbation.  Matching MOND to observations of
rotation curves of galaxies gives $g_{0}\simeq 1\times 10^{-10}\rm m\;
s^{-2}$. We further assume that $g_{0}$ is constant with time.
\subsection{Initial conditions} 

We chose a power law for the initial density contrast
  profile.  We express the mean density contrast inside a distance
$x$ from the centre of symmetry as
\begin{equation}
\bar \delta_{1}(z=0)=\delta_{0}\left(\frac{x}{x_{0}}\right)^{-s}
\end{equation}
when normalised to redshift zero ($z=0$) according to Newtonian
theory.  We choose $\delta_{0}=1.68$ so that $x_{0}$ is roughly the
Lagrangian size of the collapsed object at $z=0$. This physical
meaning for $x_{0}$ is only valid in Newtonian dynamics.  This way of
normalising the perturbation has no bearing on the final results for
MOND and is adopted here only for convenience.

The initial conditions are given at some very high redshift, $z_{1}$
as
\begin{equation}
\bar \delta_{1}(z_{1})=\frac{1}{D}\bar \delta_{1}(z=0) \; ,
\label{eq:inid} 
\end{equation}
where $D$ is the (Newtonian) linear growth factor from $z_{1}$ until
$z=0$ (e.g. Peebles 1980).  The initial velocities are set according
to the growing mode of linear theory as
\begin{equation}
v_{1}(x)=-\frac{\bar \delta_{1}(x)}{3}H_{1} x \; , \label{eq:iniv}
\end{equation}
where $H_{1}=H(z_{1})$.
The internal energy is 
\begin{equation}
u_{1}(x)=0 \; . \label{eq:iniu}
\end{equation}

\subsection{The limit of small density  perturbations}
\label{sec:small}
Here we treat the limit of small density perturbation when $\delta\ll
1$ such that the convective terms in the equations of motion can be
ignored.  We will show that in this limit the density
contrast tends to the form $\delta \propto 1/x$ independently of the
amplitude and slope of the initial contrast density $\delta_{1}\propto
x^{-s}$.  We make the following assumptions.  ($i)$ $\Omega_{\rm b}=1$
as is the case if the analysis is restricted to
  sufficiently early times, ($ii)$ the gas is initially cold and
remains cold as long as $\delta\ll 1$ , and ($iii)$ the transition to
the MOND regime is abrupt so that $g=\gN\sqrt{g_{n}/|\gN|} $ for $g\le
g_{0}$, and $g=\gN$ otherwise.

The initial mean density contrast, $\bar \delta_{1}=
\delta_{0}(x/x_{0})^{-s}$, is given at some very high redshift
$z_{1}$.  We first assume that $s>1$. In this case
$|g|\propto x\bar \delta_{1}\sim x^{1-s}$ decreases with radius, so
that at redshift $z_{0}$, there exists a distance $\xm(z)$ such that
all shells with $ x\ge \xm(z_{0})$ are in the MOND regime, while those
with $x<\xm(z_{0})$ are still in the Newtonian
  regime.  We compute below the density profile for $x\gg \xm$. The
reason for excluding shells at $x\gtsim \xm$ will become clear as we
proceed.  We assume that a shell at $x\gg \xm$ enters
the MOND regime at $z_{_{\rm M}}(x)<z_{1}$.  For $z_{1}>z>z_{_{\rm
    M}}$ all shells interior to $x$ are in the Newtonian regime and
the amplitude of the peculiar gravitation force field at $x$ is given
by
\begin{equation}
|g|=(1+z)^{2}\frac{\bar \delta(z) {G \bar M}}{ x^{2}} \; ,
\end{equation}
where $\bar \delta(z)$ is the density contrast at $z$, and ${\bar
  M}=(4\pi/3){\bar \rho} x^{3}$, $\bar \rho$ being the background
density at redshift zero.  In the Newtonian regime, we write $\bar
\delta(z)=(1+z_{1})\bar \delta_{1}/(1+z)$. Therefore,
\begin{equation}
|g|=\frac{4\pi}{3} G \bar \rho \bar \delta_{1} (1+z_1)
 (1+z) x \; .  
\end{equation} 
The transition to the MOND regime occurs at $z_{_{\rm
    M}}$ at which $|g|=g_{0}$.  Therefore,
\begin{equation}
\frac{1}{1+z_{_{\rm M}}}=
\frac{4\pi}{3}(1+z_{1})\bar \delta_{1}G \bar \rho\frac{x }{g_{0}}
\; .
\end{equation}
The density contrast in the MOND regime grows like $1/(1+z)^{2}$ for
sufficiently late times ($z\ll z_{_{\rm M}}$) (Nusser 2002).  We can
use this result if $x\gg \xm(z_{0})$ since this condition implies that
$z_{0}\ll z_{_{\rm M}}$.  The density at $z_{_{\rm M}}$ is
$(1+z_{1})\bar \delta_{1}/(1+z_{_{\rm M}})$, which must be multiplied
by $(1+z_{M})^{2}/(1+z_{0})^{2}$ to obtain the density, $\bar
\delta(z_{0})$, at $z_{0}$.  In this last step we have assumed that
the mass of matter encompassed within a sphere of
  radius $\xm$, which is Newtonian,
does not contribute significantly to $\bar \delta$ at
$x\gg\xm$.  The result is
\begin{equation}\bar \delta(x,z_{0})=\frac{3g_{0}}{4\pi G\bar \rho}
  \frac{1}{(1+z_{0})^{2}}\frac{1}{x} \; ,
\label{eq:small}
\end{equation}
for $x\gg \xm$ and $s>1$.
For $s<1$ one can show that the last expression  applies  
at $x\ll \xm$.

\section{Numerical model}
\label{sec:nmodel}
The evolution of a spherical perturbation in an otherwise uniform
Universe is followed using a one-dimensional Lagrangian hydrodynamical
code.  The code follows the evolution of of $N$ shells inside a region
of initial comoving radius $x_{\rm max}$.  At the initial time, the
shell $i+1/2$ is bounded by the $x_{i}=(i-1) \dd x$ and
$x_{i+1}=x_{i}+\dd x$ where $\dd x=x_{\rm max}/N$. The shell is then
assigned a ``mass'',
\begin{equation}
\dd \caM_{i+1/2}=\frac{4\pi}{3}\left[x_{i+1}^{3}(1+\bar \delta_{1}(x_{i+1}))
-x_{i}^{3}(1+\bar \delta_{1}(x_{i}))\right] \; ,
\end{equation}
where $\bar \delta_{1}$ is given by (\ref{eq:inid}).  This mass is
maintained constant throughout the evolution of the shell. The initial
velocities at $x=x_{i}$ are set according to (\ref{eq:iniv}).  A
leapfrog integration scheme is used and so the positions, $x_{i}^{n}$,
at the time step $t=t^{n}$ and the velocities, $v^{n+1/2}$, at
$t=t^{n-1/2}=t^{n}-\dd t^{n-1/2}/2$, are advanced as follows,
\begin{eqnarray}
\nonumber v^{n+1/2}_{i}&=&v^{n-1/2}_{i}\frac{1-H^{n}\dd t^{n}}{1+H^{n}\dd t^{n}}\\
&&-
\left[\frac{4\pi (x_i^n)^2}{(a^n)^2}\frac{{\tilde p}_{i+1/2}^n-{\tilde p}_{i-1/2}^n}
{\dd {\caM}_i}-\frac{g^{n}}{a^{n}}\right]\dd  t^{n}\; ,
\end{eqnarray}
\begin{equation}
x_{i}^{n+1}=x_{i}^{n}+v_{i}^{n+1/2}\dd t^{n+1/2} \; ,
\end{equation}
where 
\begin{equation}
\dd t^{n}=\frac{1}{2}(\dd t^{n+1/2}+\dd t^{n-1/2})\; ,
\end{equation}
\begin{equation}
\dd \caM_{i}=\frac{1}{2}(\dd \caM_{i-1/2}+\dd \caM_{i+1/2}) \; ,
\end{equation}
and $\tilde p = p+q$, where $q$ accounts for artificial viscosity.
Casting the standard expression for $q$ (e.g. Richtmyer \& Morton
1967,  Thoul \& Weinberg 1995) in terms of comoving coordinates
gives,
\begin{eqnarray}
q^{n+1}_{i+1/2}&=&-\frac{2 c_{q}(H^{n+1/2})^{2}}
{(1+\delta^{n+1}_{i+1/2})^{-1}+(1+\delta^{n}_{i+1/2})^{-1}}\\
&& |V^{n+1/2}_{i+1}-V^{n+1/2}_{i}|(V^{n+1/2}_{i+1}-V^{n+1/2}_{i}) \; ,
\end{eqnarray}
if $V^{n+1/2}_{i+1}-V^{n+1/2}_{i}<0$, and $q^{n+1}_{i+1/2}=0$ otherwise. 
Here 
\begin{equation}
V^{n+1/2}_{i}=\frac{a^{n+1/2}}{2}(x^{n+1}_{i+1}+x^{n}_i)
+\frac{a^{n+1/2}}{H^{n+1/2}} v^{n+1/2}_{i}
\end{equation}
is equal to the actual total physical velocity and we take $c_q=4$
which would spread the shock over $\sim 4 $ shells.  In the Newtonian
case the gravitational force field is
\begin{equation}
g^{n}_{i}=\frac{3 (H^{n})^{2}}{8\pi} a^n\Omega^{n}\frac{\sum_{j}\dd \caM_{j+1/2}-\frac{4\pi}{3} x_{i}^{3}}{x_{i}^{2}} \; ,
\end{equation}
where the summation is over all $j$ having $x_{j}<x_{i}$. This
expression is used in (\ref{eq:gm}) to obtain the gravitational force
field in MOND.

The density and energy are  updated, respectively,  according to 
\begin{equation}
1+\delta^{n+1}_{i+1/2}=\frac{3\dd \caM_{i+1/2}}{ 4\pi\left[(x_{i+1}^{n+1})^{3}
-(x_{i}^{n+1})^{3}\right]} \; , 
\end{equation}
and
\begin{eqnarray}
\nonumber u^{n+1}_{i+1/2}&=&u^{n}_{i+1/2}\frac{2-3H^{n+1/2}\dd t^{n+1/2}}{2+3H^{n+1/2}\dd t^{n+1/2}}\\
\nonumber &&-\left[\frac{(p^{n+1}+p^{n})}{2}+q^{n}_{i+1/2}\right]\\
&&\times \left(\frac{1}{1+\delta^{n+1}_{i+1/2}}-\frac{1}{1+\delta^{n}_{i+1/2}}          \right) \; .
\end{eqnarray}

The time-step $\dd t^{n+1/2}$ is chosen to be the minimum of
\begin{equation}
c_{d}\sqrt{\frac{a x_{i}}{g_{i}}}\; , 
\end{equation}
\begin{equation}
c_{C}\frac{x_{i}-x_{i-1}}{ \sqrt{\gamma(\gamma-1)u_{i}}}\; ,
\end{equation}
and
\begin{equation}
c_{v}\left|\frac{x_{i}-x_{i-1}}{v_{i}-v_{i-1}}\right| ,
\end{equation}
where we take $c_{d}=0.001$, $c_{C}=0.2$ and $c_{v}=0.05$ (Thoul \&
Weinberg 1995).

The boundary conditions are: $v_{i=1}=0$, and a $p_{_{N+1/2}}=0$.  We
have not included any explicit force softening in the code.  For the
number of shells we considered here the runs in all cases were
completed within a few days of CPU time on a G4 machine.

\section{Tests of the numerical scheme in the Newtonian case}
\label{sec:tests}
Since the change from Newtonian gravity to MOND is trivial (see
Eq.~\ref{eq:gm}), it suffices to confront the code with self-similar
solutions in the Newtonian case.  For an adiabatic gas, the initial
conditions given in Eq.~(\ref{eq:inid}--\ref{eq:iniv}) lead to
self-similar evolution only in a flat Universe with $\Omega_{\rm
  b}=1$. Therefore, in this section we adopt this value of
$\Omega_{\rm b}$.  Here show results from the
  simulation output at redshift, $z=0.1$.  Instead of scales
variables (e.g. temperature and distance relative to their respective
virial values) we work with physical quantities.  The temperatures are
given in degrees Kelvin, number densities in $\rm cm^{-3}$ and
distances in $\rm kpc$. The advantage of using physical quantities is
that they provide a more direct comparison with MOND where the mean
density within the virial radius is not necessarily proportional to
the background density.

We have run the code with $x_{0}=10 \; \rm Mpc$ for three values of
$s$: $s=0.25$, 2 and 2.8.  We use $N=500$ shells for the runs with
$s=0.25$ and $s=2$, and $N=100$ shells for $s=2.8$. The reason for the
smaller number of shells for $s=2.8$ is that the code is substantially
slower in this case as the time steps become extremely short at the
later stages of the evolution. All simulations were run with an
adiabatic index of $\gamma=5/3$.

In Newtonian dynamics, given the initial conditions
Eq.~(\ref{eq:inid}-\ref{eq:iniu}), the evolution of a shell can be
described qualitatively as follows. In the early stages the shell
expands until it reaches its turnaround radius at which the total
physical velocity, $ a H x+a v$, is zero.  The shell then collapses
towards the centre until it encounters the shock, transforming most of
its kinetic energy into heat and become part of the hot region.

\begin{figure}
\hspace{-0.45cm}
\includegraphics[width=3.in]{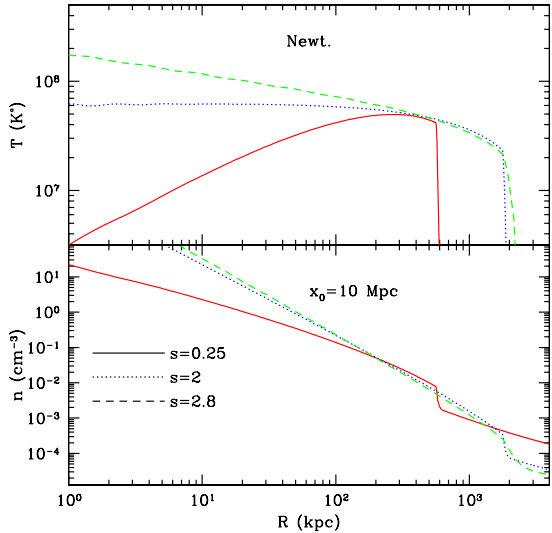}
\vspace{-0.1cm}
\caption{The temperature (top) and density (bottom) profiles 
at redshift $z=0$, obtained from the code for a 
power law initial profiles (see Eq.\ref{eq:inid}) for $x_{0}=10\; \rm  Mpc$.
The various curves correspond to different values of the power
law index, $s$, as indicated in the figure. 
The runs are for flat Universe with $\Omega_{b}=1$ and no dark matter.
All curves agree well with the self-similar solutions of Chuzhoy \& Nusser
(2000).  
\label{fig:nut}}
\end{figure}

In Fig.~\ref{fig:nut} we plot the temperature (top panel) and density
(bottom) versus the physical distance, $R$, from the output of the
simulations at redshift $z=0.1$.  The shock position, $R_{\rm sh}$, in
the various cases corresponds to the abrupt change in the density and
temperature. The transition in the shock is smoother for $s=2.8$
(dashed curve) as a result of the smaller number of shells in this
run.  The density variation across the shock for $s=0.25$ and $s=2$ is
close to $(\gamma+1)/(\gamma-1)=4$, as expected in strong shocks.
Because of the smaller number in the $s=2.8$ case, the density jump at
the shock is weaker in this case.

At each time, $t$, there is a shell that is currently at its
turnaround radius. This radius is (in physical units),
 \begin{equation}
R_{\rm ta}=x_{0} (1+z)^{-(1+1/s)}\left(\frac {5}{3}\delta_{0}\right)^{1/s}
\left(\frac{4}{3\pi}\right)^{(2/3)(1+1/s)} \; .
\label{eq:rta}
\end{equation}
We have inspected the output of the simulations at various times and
confirm that the turaround radii agree very well with (\ref{eq:rta}).
For $s=0.25$ and $s=2$, Chuzhoy \& Nusser (2000) (hereafter CN2000)
provided full self-similar solutions (see their figure~2).  The shapes
of our profiles match extremely well the solutions of CN2000.  We also
checked if the shock radius relative to the turnaround radius agrees
with the self-similar solutions.  At redshift $z=0.1$ we find, from
(\ref{eq:rta}), that $R_{\rm ta}=6.155 \;\rm Mpc$ and $21.9 ~\rm Mpc$
for $s=2$ and $0.25$, respectively.  Inferring the shock radii,
$R_{\rm sh}$, from Fig.~\ref{fig:nut} we get $R_{\rm sh}/R_{\rm
  ta}\approx 1.8/6.155=0.29$ for $s=2$ and $0.57/21.9=0.026$ for
$s=0.25$. These ratios agree very well with those inferred from figure
2 of CN2000.

For $s=2.8$ and $s=2$, the asymptotic logarithmic
slopes near the origin agree very well with the analytic slopes
presented in Table 1 of CN2000. For $s=0.25$, neither our curves nor
those in figure 2 of CN2000 behave according to the asymptotic
exponents in that Table.  This is because the analytic exponents in
the $s=0.25$ case are realized at extremely small fractions of the
shock radius (see CN2000).

\section{Results with  MOND}
\label{sec:results}

All results here will be presented for $g_{0}=1\times 10^{-10}\rm m\;
s^{-2}$ and $\Omega_{\rm b}=0.044$.  This value of $\Omega_{\rm b}$ is
consistent with the observations of the deuterium abundance relative
to hydrogen in Ly$\alpha$ clouds at high redshift (Kirkman et~al.
2000). It is also consistent with the ratio of the first to second
peak heights in the angular power spectrum of the CMB anisotropies
(e.g. Spergel et~al. 2003). We further take $h=0.7$ for the present
Hubble constant (in units of $100\; \rm km \; s^{-1} Mpc^{-1}$).  We
work with a vanishing cosmological constant since otherwise the age of
the Universe will be unreasonable large.  For the cosmological
parameters of choice, the age of the Universe $\approx 14\rm Gyr$.

\begin{figure}
\hspace{-0.45cm}
\includegraphics[width=3.in]{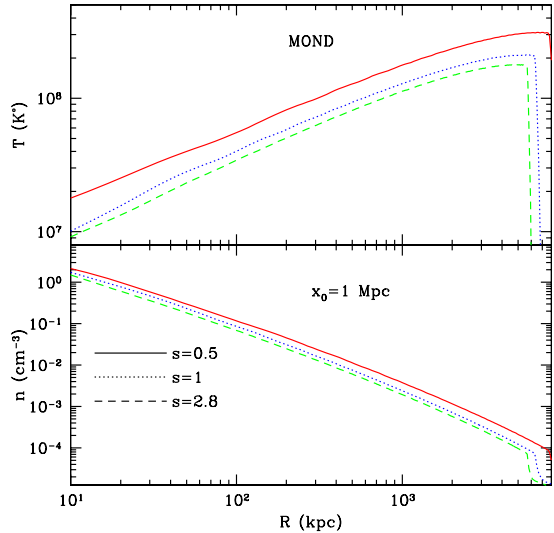}
\vspace{-0.1cm}
\caption{The temperature (top) and the density (bottom) profiles in MOND 
for $s=0.5$, 1 and 2.8, as indicated in the figure. 
The normalisation is such that the mean value of the density 
within a comoving distance $x_{0}=1 \rm Mpc$ is the same 
at the initial time for all three values of $s$. 
\label{fig:alpha}}
\end{figure}

\begin{figure}
\hspace{-0.45cm}
\includegraphics[width=3.in]{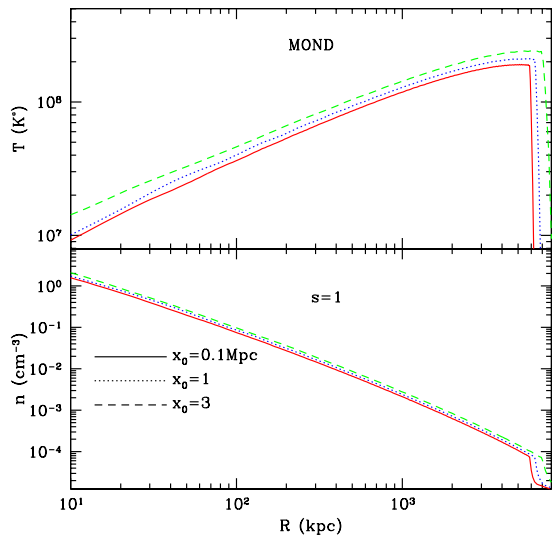}
\vspace{-0.1cm}
\caption{Profiles in MOND for $s=1$, but for three values of the 
normalisation as indicated in the figure by the parameter $x_{0}$.
\label{fig:beta}}
\end{figure}

The initial redshifts in all the runs is chosen sufficiently high so
that all shells in the simulations are still Newtonian.
 We investigated various values for the parameters
$s$ and $x_{0}$ (see Eq.~\ref{eq:inid}).  In Fig.~\ref{fig:alpha} and
Fig.~\ref{fig:beta} we show the temperature and the local baryon
number density (in $\rm cm^{-3}$) versus the physical distance from
the centre, $R$, for three different values of $s$, but the same
$x_{0}$.  The normalisation of the initial density profile is such
that the average density contrast within a comoving distance $x_{0}$
is the same for all three values of $s$.  There is a striking
similarity in the shapes of the temperature and density profiles,
respectively.  This is in agreement with the analytic considerations
in section \S~\ref{sec:small}.  The profiles are also insensitive to
$x_{0}$ as seen in Fig.~\ref{fig:beta} which shows results for $s=1$
but different $x_{0}$. Even a change of a factor of 30 in $x_{0}$
fails to amount to any significant variation in the shape.  The
overall amplitude of both the temperature and the density profiles is
also insensitive to $x_{0}$.  The density and the temperature show,
respectively, $\sim R^{-1.5}$ and $\sim R^{0.5}$, dependence on
radius.

\begin{figure}
\hspace{-0.45cm}
\includegraphics[width=3.in]{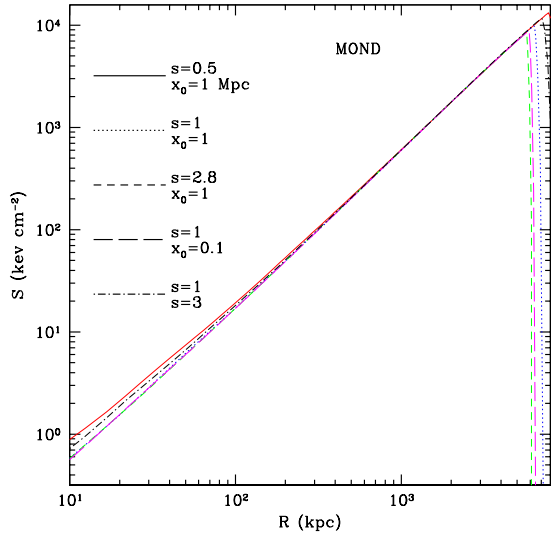}
\vspace{-0.1cm}
\caption{Entropy profiles  in MOND at $z=0.1$ for 
various values of $s$ and $x_{0}$.
The slope, $\dd \ln S/\dd \ln R$,  at $R\gtsim 10^{3}\rm kpc$ is 1.3, approximately.
\label{fig:eta}}
\end{figure}

The ``entropy'', $S=kT/n^{2/3}$, is plotted in Fig.~\ref{fig:eta}.
The entropy profiles corresponding to all the
  simulation outputs presented in the previous two figures are shown.
In the shock heated regions the curves are practically
indistinguishable at radii $R\gtsim 200\rm kpc$.  At these radii,
$S\propto r^{1.5}$.

\begin{figure}
\hspace{-0.45cm}
\includegraphics[width=3.in]{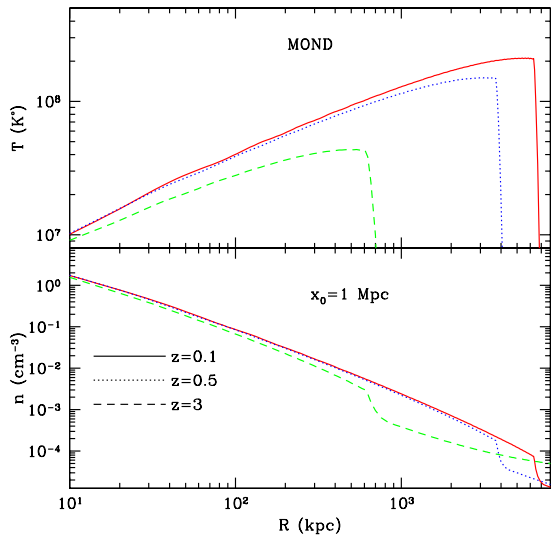}
\vspace{-0.1cm}
\caption{Profiles in MOND for $s=1$ and $x_{0}$ from outputs at 
three different redshifts, $z$, as indicated in the figure.  
\label{fig:betaz}}
\end{figure}

The redshift evolution for the case $s=1$ and $x_{0}=1\; \rm Mpc$ is
explored in Fig.~\ref{fig:betaz} showing profiles at three different
redshifts.  There is an increase by a factor of about
  1.8 in the shock location between $z=0.5$ and $z=0.1$, but the
profiles at these two redshifts have similar shapes.  Over the
distance scales in the plot the temperature shows a significant change
from $z=3$ until the present. But the overall shape at $z=3$ is
similar to the outer portion of the low redshift profiles. This
indicates that the shape is preserved with redshift.  The density
profiles inside the shocked region at the three redshifts are
remarkably similar, indicating that the system is in hydrostatic
equilibrium.

\begin{figure}
\hspace{-0.45cm}
\includegraphics[width=3.in]{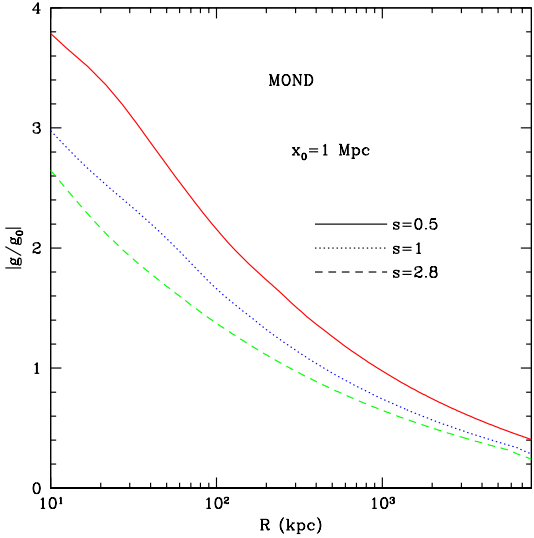}
\vspace{-0.1cm}
\caption{The ratio of the absolute value of the peculiar gravitational
field relative to $g_{0}=1\times 10^{-10}\rm m s^{-2}$ as a function 
of radius for  runs with the same normalisation and three
different values of $s$.
\label{fig:gamma}}
\end{figure}

At early epochs near the initial times, perturbations evolves
according to Newtonian dynamics and then, as $|g|$ decreases, they
succumb to MOND.  This is as long as the perturbations are small.
Fig.~\ref{fig:gamma} demonstrates that in collapsed regions the
perturbations become Newtonian again.  The figure shows $|g|$ relative
to $g_{0}$ as a function of $R$.  In all cases $|g/g_{0}|>1$ in the
inner regions.

The insensitivity of the profiles to the initial conditions is
consistent with our findings in \S\ref{sec:small}.  A perturbation
enters the MOND regime when the density fluctuations are still small.
During this stage of the evolution they acquire the form
(\ref{eq:small}) independently of the initial density profile, as
shown in \S\ref{sec:small}. Later evolution, when the density
fluctuations are large, modified the form given in \S\ref{sec:small},
but can not restore the dependence on the initial conditions.  It is
interesting that the evolved form of the density profile in MOND is
$\sim r^{1.5}$ which is the same as the one obtained in the
self-similar collapse of an initial $1/x$ perturbation in Newtonian
dynamics (e.g. CN2000).

\section{The observational situation}   
\label{sec:obs}
\begin{figure}
\hspace{-0.45cm}
\includegraphics[width=3.in]{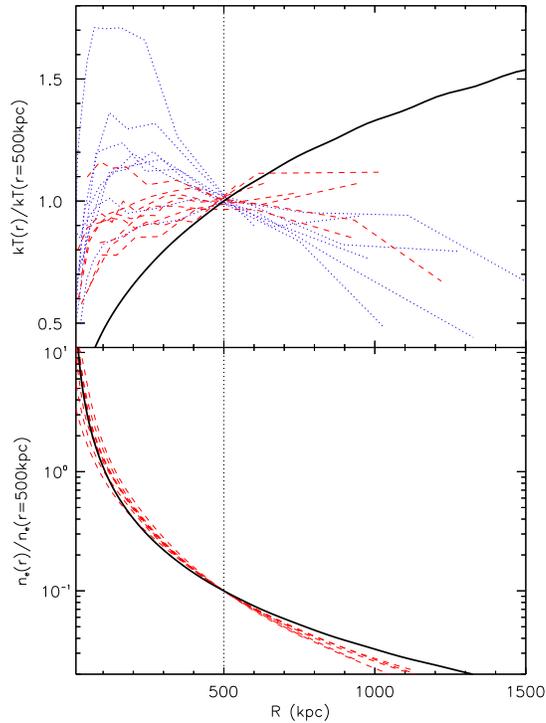}
\vspace{-0.1cm}
\caption{ {\it Top}: MOND simulated temperature profiles (thick black line) compared to the individual profiles 
  of clusters observed with XMM-\emph{Newton} (red
  dashed lines) and {\it Chandra} (blue dotted
  lines). {\it Bottom panel}: MOND simulated density profiles (thick
  black line) shown with respect to the observed density profiles of nearby
  clusters by the XMM-\emph{Newton} satellite (red dashed lines). In
  both panels the profiles have been normalised according to their
  respective values at $r=500$~kpc as marked radius by dotted vertical
  line.
  \label{fig:ktden}}
\end{figure}

We confront the temperature and density profiles of MOND with recent
X-ray observations of galaxies clusters.  We use the recently
published temperature profiles by Pointecouteau et~al. (2005) obtained
for a sample of 10 nearby ($z<0.15$) clusters with the
XMM-\emph{Newton} satellite. To give a fair idea of the variation in
shape of the observed temperature profiles, we also add the
temperature profiles of eleven nearby ($z<0.23$) clusters obtained by
Vikhlinin et~al. (2004) with the {\it Chandra} satellite.  Both
samples cover a wide range of temperatures from 2 to 9~keV.  We scale
each temperature profile to its value at $500$~kpc (i.e
$kT(r=500\textrm{~kpc})$) in order to compare them with our simulated
profiles. To avoid any extrapolation, we kept only the 16 clusters
which were observed beyond 500~kpc (i.e 8 XMM-\emph{Newton} and 8
Chandra clusters). The scaled temperature profiles are shown in
Fig.~\ref{fig:ktden} (top panel).

We also compare the observed density profiles with MOND's profiles.
The observed profiles correspond to the parameterised form derived by
fitting the X-ray surface brightness profile of each cluster. For this
comparison, we only use the XMM-\emph{Newton} profiles. The best fit
models for the density profiles of the ten clusters from Pointecouteau
et~al. (2005) are plotted together with the MOND density profile in
Fig.~\ref{fig:ktden} (bottom panel). All profiles are scaled to their
respective values at $500\; \rm kpc$.

All of the observed clusters are identified as relaxed and show small
deviations from sphericity.  The temperature drop towards the centre
in the observed profiles is the result of cooling. Since cooling is
not included in our simulations, we restrict the comparison to
$R\gtsim 200 \; \rm kpc$.

For better visualisation, all profiles are scaled to their respective
values at $R=500\;\rm kpc$.  At that radius, the minimum, maximum,
average, and median values of the observed temperature are (in
$10^{7}{\rm K}$), 1.9, 11.9, 6.6, and 6.8, respectively. The corresponding values for
the baryon number density are (in $10^{-4}\rm cm^{-3}$), 4.4, 20.6,
13.2, and 18.2, respectively.  At this radius MOND yields a
temperature range consistent with the observations.  However, among
the MOND runs of all values of $s$ and $x_{0}$ that we shown, the
minimum value number density at $500\; \rm kpc$ is MOND density in
MOND is $0.07 \; \rm cm^{-3}$. This is substantially higher that the
maximum observed number density of $0.0026\; \rm cm^{-3}$.  More
importantly, the temperature profile, $\sim r^{0.5}$, in MOND is, in
clear contrast to the observed profiles, either flatten or show a mild
decline in the outer regions.  The entropy profile in MOND has a
logarithmic slope of $\sim 1.5$, as compared to $0.94\pm 0.14$ and
$0.95\pm 0.02$ found in the observations (Pratt \& Arnaud 2005 and
Piffaretti \etal\ 2005, respectively). This is another manifestation
of the mismatch between temperature profiles in MOND and the
observations.


\section{Summary and discussion}
\label{sec:sum}

In this paper we aimed at: (1) studying the density and temperature
profiles of spherically symmetric perturbations evolved under MONDian
gravity, and (2) comparing the predicted profiles with recent X-ray
observations of the intracluster medium (ICM).  A key finding here is
that the profiles in MOND acquire a nearly universal shape
independently of the initial fluctuations.  This is in contrast to the
Newtonian theory where the profiles are sensitive to the shape of the
initial profiles.  The shock radius in MOND is nearly independent of
the amplitude of the initial perturbation while in the Newtonian case
it is proportional to the amplitude (see Eq.~\ref{eq:rta}).

The insensitivity to the initial density profile, leaves very little
leeway in matching MOND predictions to observed properties of the ICM.
In Newtonian dynamics, one can, to a great extent, resort to varying
initial conditions to try to match the observations. In MOND one has
to depend, almost entirely, on non-gravitational physical processes.
One such process may be thermal conduction.  The thermal conduction
time scale is estimated as $R_{\rm c}\sim (2\kappa T t/5nkT)^{1/2}$.
Assuming a thermal conductivity, $\kappa$, of 0.2 times the Spitzer
value $\kappa_{\rm Sp}=9.2 10^{-7}T^{5/2}\rm erg\; s^{-1}\; K^{-1} \;
cm ^{-2}$ (Spitzer 1962), we obtained $R_{\rm c}\sim 200\; \rm kpc$ for
$t=14\; \rm Gyr$, $n=0.001\rm cm^{-3}$ (see Fig.~\ref{fig:alpha} at
$R=10^{3}\; \rm kpc$), and $T=5\times 10^{7}\; \rm K$.  Therefore,
thermal conduction can help, especially for the highest temperature
clusters of $T\sim 10^{8}\; \rm K$. For $T\ltsim 5 \times 10^{7}\rm \;
K$ conduction seems inefficient at producing the desired temperature
profile. Note that we have used the cosmic time $t=t_{\rm H}\approx
14\rm \; Gyr$ to estimate $R_{c}$. In the outer regions one should
actually take $t_{\rm H}-t_{\rm collapse}$ where $t_{\rm collapse}$ is
the time at which these regions collapsed and shock-heated. According
to Fig.~\ref{fig:betaz} the shock radius have grown from $600 \rm \;
kpc$ at $z=3$ to $7000\rm \; kpc$ at $z=0$. The time interval between
$z=3$ and $z=0$ is $t_{\rm H}-t_{\rm collapse}=0.25 t_{\rm H}$ which
reduces $R_{\rm c}$ by a factor of $2$.  Non-gravitational heating of
the ICM can also be invoked to alter the profiles, but it is hard to
imagine a realistic mechanism that will flatten the temperature in the
outer regions (e.g. Borgani et~al. 2005).

Our findings as well as those of previous authors, pause 
some puzzles for MOND  to overcome. While Newtonian dynamics
has its own  problems in matching the cluster data,
the lack of dependence on the initial conditions, makes MOND
harder to reconcile with these data.

\section{Acknowledgements} 
We are deeply indebted to Alexey Vikhlinin for kindly providing his
{\it Chandra} temperature profiles.  AN thanks Tom Theuns for advice
on the numerical code.  AN is also grateful to Arif Babul and Scott
Kay for many stimulating conversations.  EP aknowledges the financial
support of the Leverhulme Trust (UK).

  \protect



\end{document}